\documentclass[runningheads]{llncs}

\usepackage{comment}

\usepackage[utf8]{inputenc} 
\usepackage[T1]{fontenc}    
\usepackage{hyperref}       
\usepackage{url}            
\usepackage{booktabs}       
\usepackage{amsfonts}       
\usepackage{nicefrac}       
\usepackage{microtype}      
\usepackage{lipsum}
\usepackage[dvipdfmx]{graphicx}
\usepackage{algorithmic}
\usepackage{algorithm}
\usepackage{multirow}
\usepackage{wrapfig}
\usepackage{listings}
\usepackage{enumitem}
\usepackage{graphicx}
\usepackage{amsmath,amssymb,amsfonts}

\usepackage{caption}
\usepackage{subcaption}
\usepackage{threeparttable}

\usepackage{xcolor}
\usepackage{pifont}
\newcommand{\cmark}{{\color{green}\ding{51}}}%
\newcommand{\xmark}{{\color{red}\ding{55}}}%

\begin{document}
\title{TFHE-SBC: Software Designs for Fully Homomorphic Encryption over the Torus on Single Board Computers}
\titlerunning{TFHE-SBC: Software Designs for TFHE on SBCs}
%
\author{Marin Matsumoto\inst{1}\and
Ai Nozaki\inst{2}\and
Hideki Takase\inst{2}\and
Masato Oguchi\inst{1}}
\authorrunning{M. Matsumoto et al.}
%
\institute{Ochanomizu University, Tokyo, Japan\\ \email{marin@ogl.is.ocha.ac.jp, oguchi@is.ocha.ac.jp} \and
The University of Tokyo, Tokyo, Japan\\
\email{\{nozaki,takasehideki\}@hal.ipc.i.u-tokyo.ac.jp}}
\maketitle              
\begin{abstract}
Fully homomorphic encryption (FHE) is a technique that enables statistical processing and machine learning while protecting data, including sensitive information collected by single board computers (SBCs), on a cloud server. Among FHE schemes, the TFHE scheme is capable of homomorphic NAND operations and, unlike other FHE schemes, can perform various operations such as minimum, maximum, and comparison. However, TFHE requires Torus Learning With Error (TLWE) encryption, which encrypts one bit at a time, leading to less efficient encryption and larger ciphertext size compared to other schemes. Additionally, SBCs have a limited number of hardware accelerators compared to servers, making it challenging to achieve the same level of optimization as on servers. In this study, we propose a novel SBC-specific design, \textsf{TFHE-SBC}, to accelerate client-side TFHE operations and enhance communication and energy efficiency. Experimental results demonstrate that \textsf{TFHE-SBC} encryption is up to 2486 times faster, improves communication efficiency by 512 times, and achieves 12 to 2004 times greater energy efficiency than the state-of-the-art.
\end{abstract}

\section{Introduction}
Single board computers (SBCs), such as the Raspberry Pi, can collect various data from temperature, humidity, air pressure, and illumination sensors at low cost and with high energy efficiency. 
When sensitive information is included in the data collected by SBCs, fully homomorphic encryption (FHE) becomes crucial for securely analyzing the SBC data on a cloud server.
FHE is a cryptographic technique that enables operations on encrypted data without decrypting it.
The main FHE schemes are BGV~\cite{brakerski2014leveled}, BFV~\cite{fan2012somewhat}, CKKS~\cite{cheon2017homomorphic}, and TFHE~\cite{chillotti2020tfhe}, based on the Learning with Errors (LWE) or Ring LWE (RLWE) problem.
Each of these schemes supports different types of plaintext (integers, floats, scalars, vectors, etc.) and different homomorphic operations.
Among the main FHE schemes, only the TFHE scheme can perform any operation, including minimum, maximum, and comparison operations.
Therefore, the TFHE scheme supports a wide range of applications such as query execution on encrypted databases~\cite{ren2022heda,bian2023he3db,zhang2024arcedb} and encrypted machine learning~\cite{benamira2023tt,chillotti2021programmable,lou2019she,lou2020glyph}.

However, the computationally intensive nature of homomorphic encryption operations has largely prevented TFHE from being adopted in the IoT domain. 
For example, a simple TFHE encryption using the TFHEpp~\cite{TFHEpp} library requires a minimum of 4.2 MiB of memory allocation when encrypting only 1 KiB with a commonly used parameter set.
This characteristic arises from the fact that the TFHE scheme processes Torus LWE (TLWE) encryption, which encrypts data one bit at a time.
TLWE encryption results in inefficient encryption and large ciphertext sizes and is inconvenient for SBCs with limited computational resources.
Additionally, SBCs have a limited number of hardware accelerators compared to servers, making it difficult to achieve the same level of optimization as on servers.
To make such an SBC-unfriendly scheme feasible in IoT systems, we identify the following three requirements:
\begin{itemize}
    \item[\textbf{R1}:] Computation efficiency on SBCs without special accelerators and on the server.
    \item[\textbf{R2}:] Communication efficiency between SBCs and the server.
    \item[\textbf{R3}:] Energy-efficient encryption solutions for SBCs.
\end{itemize}

\begin{figure}[t]
    \centering
    \includegraphics[width=.7\columnwidth]{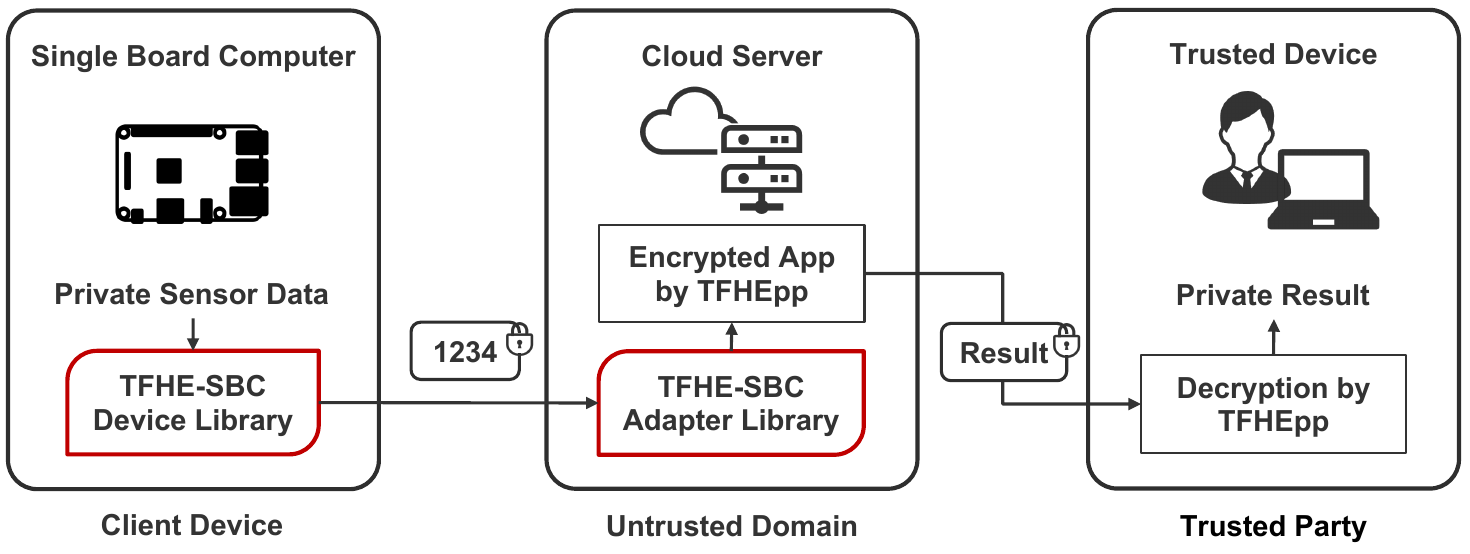}
    \caption{End-to-end TFHE deployment solution by \textsf{TFHE-SBC}.}
    \label{fig:tfhe-sbc-overview}
    \vspace{-5pt}
\end{figure}

To address the above requirements, we propose \textsf{TFHE-SBC}, the first TFHE framework targeted for SBCs that can simultaneously achieve efficient encryption and communication.
Existing implementations of the TFHE scheme use inefficient TLWE encryption, but \textsf{TFHE-SBC} differs from existing implementations in that it uses TRLWE encryption.
Torus RLWE (TRLWE) can process multiple bits simultaneously, resulting in better processing speed and smaller ciphertext size compared to TLWE.
As shown in Figure \ref{fig:tfhe-sbc-overview}, it consists of a device library that encrypts multiple bits at once using TRLWE encryption on the client side and an adapter library that converts them into the TLWE ciphertext needed for the TFHE application on the server side.
We also propose a modified random number generation algorithm and a memory reuse strategy to improve the efficiency of TRLWE encryption on the limited computational resources of the SBC.
The main contributions of this work are summarized as follows.
\vspace{-5pt}
\begin{itemize}
\item To the best of our knowledge, we are the first to employ ciphertext type switching, which converts TLWE to TRLWE, to achieve efficient encryption and communication efficiency on SBCs.
This solution also improves energy efficiency.

\item We identify the specific memory and performance challenges related to enabling TFHE on SBCs and discuss several techniques to overcome them. 
In particular, we show how to design random number generation algorithms without environment-dependent optimization.
We further investigated the feasibility of accelerating polynomial multiplication with the VideoCore IV (VC4) GPU installed in the Raspberry Pi Zero 2W and experimentally showed that there are accuracy issues.
\item We provide encryption time, communication costs, memory usage, and energy consumption results for a variety of configurations enabled by \textsf{TFHE-SBC} for a parameter set that enables insightful analytics.
As shown in Table \ref{tab:ex-tlwe}, we achieved 15 to 2486 times faster encryption, up to 512 times more efficient communication, up to 12.8 times more memory efficiency, and 12 to 2004 times more energy efficiency than existing implementations.
\end{itemize}
\vspace{-5pt}

\begin{table*}[t]
 \caption{Benchmark results of client-side operations for the TFHE scheme on Raspberry Pi Zero 2W. \textsf{TFHE-SBC} outperforms other methods in encryption time, communication cost, memory usage, and energy consumption.}
    \centering
    \resizebox{\textwidth}{!}{
    \begin{tabular}{|c|c||cc||cc||cc||cc|}
    \hline
         Plaintext&Methods& \multicolumn{2}{c||}{\textbf{Encryption Time [ms]}}& \multicolumn{2}{c||}{\textbf{Ciphertext [KiB]}}&\multicolumn{2}{c||}{\textbf{RAM [KiB]}}&\multicolumn{2}{c|}{\textbf{Energy [mJ]}}\\\hline\hline
        \multirow{3}{*}{256 bit} & TLWE by TFHEpp &1922.6 & 620 $\times$& 1024&128 $\times$& 1217.5& 3.6 $\times$ &2499.4&499.9 $\times$\\
        &TRLWE by TFHEpp& 46.1 & 14.9 $\times$& 8& 1 $\times$&332.2&0.99 $\times$&64.5&12.9 $\times$\\
        &\textsf{TFHE-SBC}& \textbf{3.1} & \textbf{1 $\times$}& \textbf{8}& \textbf{1 $\times$}&\textbf{336.2}& \textbf{1 $\times$}&\textbf{5.0}&\textbf{1 $\times$}\\\hline\hline
         \multirow{3}{*}{1024 bit} &TLWE by TFHEpp& 7705.6&2486 $\times$ &4096&512 $\times$&4297.7&12.8 $\times$&10017.3&2003.5 $\times$\\
          &TRLWE by TFHEpp&46.1&14.9 $\times$& 8&1 $\times$&332.2&0.99 $\times$&64.5&12.9 $\times$\\
          &\textsf{TFHE-SBC}& \textbf{3.1} & \textbf{1 $\times$}& \textbf{8}&\textbf{1 $\times$}&\textbf{336.2}&\textbf{1 $\times$}&\textbf{5.0}&\textbf{1 $\times$}\\\hline\hline
         \multirow{3}{*}{4096 bit} & TLWE by TFHEpp &30781.2&2386 $\times$&16384&512 $\times$&16599.04& 46.1 $\times$&40015.6&1914.6 $\times$\\
          &TRLWE by TFHEpp&183.7&14.2 $\times$& 32&1 $\times$&356.2&0.99 $\times$&257.1&12.3 $\times$\\
          &\textsf{TFHE-SBC}& \textbf{12.9}& \textbf{1 $\times$}& \textbf{32}&\textbf{1 $\times$}&\textbf{360.2}& \textbf{1 $\times$}&\textbf{20.9}&\textbf{1 $\times$}\\\hline\hline
          \multirow{3}{*}{8192 bit} & TLWE by TFHEpp &61587.5&2406 $\times$ &32768&512 $\times$& 33003.5&84.1 $\times$&80063.8&1929.2 $\times$\\
          &TRLWE by TFHEpp&370.8&14.5$\times$&64&1 $\times$&388.2&0.99 $\times$&519.1 & 12.5 $\times$\\
          &\textsf{TFHE-SBC}& \textbf{25.6} & \textbf{1 $\times$}& \textbf{64}&\textbf{1 $\times$}&\textbf{392.2} & \textbf{1 $\times$ }&\textbf{41.5}& \textbf{1 $\times$}\\\hline
    \end{tabular}}
    \label{tab:ex-tlwe}
    \vspace{-5pt}
\end{table*}

\section{Preliminaries}

\noindent\textbf{Notations.}
{This work requires understanding of} several mathematical objects.
$\mathcal{R} = \mathbb{Z}[X]/(X^N +1)$ {denotes} the ring of integer polynomials modulo the cyclotomic polynomial $X^N + 1$, {where $N$ is} a power of 2.
$\mathcal{R}_q = (\mathbb{Z}/q\mathbb{Z})[X]/(X^N +1)$ {represents} the same ring {$\mathcal{R}$}, but {with} coefficients modulo $q$. 
{We} often {denote} $\mathbb{Z}/q\mathbb{Z}$ as $\mathbb{Z}_q$.
The letter ``T'' in TFHE \cite{chillotti2020tfhe} refers to the real torus $\mathbb{T}:=\mathbb{R}/\mathbb{Z}$, which is the set $[0, 1)$ of real numbers modulo $1$.

\subsection{General LWE}
\label{sec:glwe}
The security of {HE schemes} is based on lattice problems such as {Learning With Errors} (LWE) and its variants, {including Ring LWE} (RLWE).
BGV, BFV, CKKS, and TFHE schemes {all build upon} RLWE or LWE ciphertexts.
{We use} a generalization that {covers} both {variants}, called General LWE {(GLWE)}.

Let $p$ and $q$ be two positive integers {with} $p\leq q$, and define $\Delta = q/p$. 
In TFHE, $p$ and $q$ are {typically} chosen {as} powers of two. 
{We refer to} $q$ {as the} ciphertext modulus, $p$ {as the} plaintext modulus, and $\Delta$ {as the} scaling factor.

\noindent
\textbf{Key Generation.} {The} secret key $\vec{S} = (S_0, \ldots, S_{k-1}) \in \mathcal{R}^k$ {consists of} $k$ polynomials of degree {less} than $N$, sampled from a uniform binary distribution.

\noindent
\textbf{Encryption.} 
A GLWE ciphertext encrypting message $M$ under {secret} key $\vec{S}$ is a tuple:
\begin{align*}
(A_0, \ldots, A_{k-1}, B) \in GLWE_{\vec{S}, \sigma}(\Delta M) \subseteq \mathcal{R}_q^{k+1}\\
\text{where}\; 
    B = \sum_{i=0}^{k-1} A_i \cdot S_i + \Delta M + E \in \mathcal{R}_q
\end{align*}
To encrypt message $M \in \mathcal{R}_p$, we sample a uniformly random mask $A_i \in \mathcal{R}_q$ and {error term} $E\in \mathcal{R}_q$ with coefficients {drawn} from a Gaussian distribution $\chi_{\sigma}$.

Since FHE encrypts by adding noise to plaintext, noise accumulates in the ciphertext as homomorphic operations are {performed}.
If noise exceeds a threshold, correct decryption becomes impossible.
Bootstrapping~\cite{gentry2009fully} is a special operation that reduces ciphertext noise.

\noindent
\textbf{Construction of LWE and RLWE.}
\label{sec:lwe-and-rlwe}
LWE is {instantiated from} GLWE by setting $k = n \in \mathbb{Z}$ and $N = 1$.
RLWE is {instantiated from} GLWE by setting $k=1$ and $N$ as a power of 2.
LWE encrypts {scalar} messages, while RLWE encrypts {polynomial} messages.
LWE encryption involves the inner product of a uniformly random mask and the secret key,
whereas RLWE encryption involves polynomial multiplication of polynomial $A$ and secret key $S$.

\subsection{TFHE Scheme}
The TFHE scheme {operates with} three {ciphertext} forms: TLWE, TRLWE, and Torus Ring Gentry Sahai Waters (TRGSW).
TLWE and TRLWE are constructions of LWE and RLWE over the torus (the set of fractional parts of real numbers), respectively.
A TLWE ciphertext {represents} a Boolean value, while a TRLWE ciphertext {represents} a vector of $N$ Boolean values.
TRGSW contains multiple TRLWE ciphertexts and {represents} the bootstrapping key required for the TFHE bootstrapping process.
The {fundamental} ciphertext format in TFHE is TLWE.

\noindent\textbf{Sample Extraction.}
\label{sec:sample-ext}
Sample Extraction~\cite{chillotti2020tfhe} is a component of the TFHE bootstrapping process.
This operation takes a TRLWE ciphertext encrypting a polynomial message and extracts the encryption of a {single} coefficient as a TLWE ciphertext.
The operation {maintains the noise level} and consists of {strategically copying} coefficients from the TRLWE ciphertext {to construct} the output TLWE ciphertext.
Algorithm \ref{alg:convert} {details} the implementation.

\subsection{Multiplying Polynomials via the Discrete Weighted Transform}
Polynomial multiplication is {essential} for (T)RLWE encryption and the TFHE bootstrapping process.
{We describe} polynomial multiplication {using} both the ordinary discrete Fourier transform (DFT) and the discrete weighted transform (DWT).
Let us define polynomials of degree {at most} $N-1$ as $f(x)=f_0+f_1x\cdots+f_{N-1}x^{N-1}$ and $g(x)=g_0+g_1x\cdots+g_{N-1}x^{N-1}$,
where $f$ and $g$ are assumed to be periodic such that $f(x+N)=f(x)$ and $g(x+N)=g(x)$.

The convolution $f*g$ for discrete values is defined as $(f*g)(x)=\sum_{n=0}^{N-1}f(n)g(x-n)$ and is equivalent to polynomial multiplication modulo $x^N-1$.
The {DFT} of polynomial $f$ is defined as
$F(t)=\sum_{x=0}^{N-1} f(x)e^{-i\frac{2\pi tx}{N}}$.
The inverse discrete Fourier transform (IDFT) is defined as $f(x)=\frac1N\sum_{t=0}^{N-1} F(t)e^{i\frac{2\pi tx}{N}}$,
where $e^{i\frac{2\pi tx}{N}}$ {are} called twiddle factors.

To {compute} $f*g$, we {calculate} the DFT of both $f$ and $g$, multiply their coefficients {pointwise}, and apply the IDFT.
The computational complexity of DFT is $\mathcal{O}(N^2)$, but for $N=2^k$, fast Fourier transform (FFT) and inverse FFT (IFFT) reduce this to $\mathcal{O}(N\log N)$.

TRLWE requires polynomial multiplication modulo $x^N+1$ (negacyclic convolution), {necessitating} the use of DWT.
The polynomial after DWT is {denoted} $F'(t)$, and {DWT/IDWT are performed} by multiplying by weight $w_x=e^{i\frac{2\pi tx}{2N}}$ as follows:
\begin{align*}
F'(t) = \sum_{x=0}^{N-1}w_x f(x)e^{-i\frac{2\pi tx}{N}}, \quad
    f(x)=\frac{1}{w_x N}\sum_{t=0}^{N-1} F'(t)e^{i\frac{2\pi tx}{N}}
\end{align*}

\section{Related Work}
In this section, we present research on software and hardware implementations of client-side FHE operations (See Table \ref{tab:summary-related-work} for a summary) and Transciphering, which reduces the amount of communication between client and server. 

SEAL-Embedded~\cite{natarajan2021seal} has accelerated CKKS encoding and encryption processes on embedded devices such as Azure Sphere Cortex-A7, Nordic nRF52840 Cortex-M4.
The client-side operations of the CKKS scheme consist of encoding and RLWE encryption processes, requiring the implementation of FFT for encoding and polynomial multiplication and random number generation for RLWE encryption.
In SEAL-Embedded, each layer is optimized to implement a different random number generation algorithm, faster polynomial multiplication using Number Theoretic Transform (NTT), and memory reuse.
As a software-based solution, \textsf{TFHE-SBC} makes the novel contribution of being the first TFHE framework that enables both efficient encryption and reduced communication costs for IoT devices, particularly SBCs. 

\cite{wang2024compact,di2023vlsi,van2021practical,krieger2023aloha} are FPGA-based accelerators for FHE.
RACE~\cite{azad2022race} and RISE~\cite{azad2023rise} implement encryption of CKKS schemes, including NTT on ASICs, but the encoding process before encryption is implemented in software.
Aloha-HE~\cite{krieger2023aloha} implements a faster encoding and encryption process for the CKKS scheme on FPGA. 
The implementation includes an FFT unit for encoding, an NTT unit for encryption, and a random number generator unit.
The encoding and decoding speeds of CAEA~\cite{wang2024compact} are both 2.01 times faster, and the encryption and decryption speeds of CAEA are 1.13 times and 3.04 times faster than Aloha-HE, respectively.
However, these ASIC implementations, while powerful, lack the flexibility in adapting to evolving requirements. 
More significantly, previous work has been limited to the CKKS scheme, leaving a substantial gap in client-side acceleration for the TFHE scheme, which offers more flexible application.

In the domain of communication efficiency, Transciphering~\cite{gentry2012homomorphic,aharoni2023e2e,wei2023fregata,trama2023last,bon2023optimized} has been proposed as a method to reduce communication overhead between client and server compared to using homomorphic encryption schemes alone. 
A client does not need to encrypt all its messages using an HE algorithm (except the symmetric key, such as the AES key); all the messages can be encrypted using only a symmetric cipher. 
The server obtains the message encrypted with HE by homomorphically evaluating the AES decryption process.
On the other hand, this conversion on the server requires at least around 6 seconds~\cite{gentry2012homomorphic} under the BGV scheme and around 9 seconds~\cite{wei2023fregata} under the TFHE scheme.
Therefore, although Transciphering reduces the amount of client-server communication, it tends to increase the overall latency of the application.
Our framework makes a substantial contribution to the communication efficiency domain by achieving reduced communication costs with significantly lower server computational overhead compared with Transciphering.

\begin{table*}[tb]\centering
    \begin{threeparttable}
	\caption{Related work on homomorphic encryption \emph{on the client}.
	}\label{tab:summary-related-work}
	\scriptsize
	\begin{tabular}{l|c|c|c|c|c|c|c||c}\toprule
		                       &\;\;\cite{wang2024compact}\;\; &\;\cite{di2023vlsi}\; &  \;\cite{van2021practical}\; &  \;\cite{azad2022race}\; &  \;\cite{azad2023rise}\; &  \;\;\cite{krieger2023aloha}\;\; &  \;\cite{natarajan2021seal}\; & \;Ours\;      \\\midrule
		Support TFHE scheme&   \xmark       & \xmark        & \xmark        & \xmark       & \xmark     & \xmark         & \xmark     & \cmark\\
		Software acceleration       & \xmark\tnote{1,2}       & \xmark\tnote{1}        & \xmark\tnote{1}        & \xmark\tnote{2}       & \xmark\tnote{2}     & \xmark\tnote{1}         & \cmark     & \cmark    \\
		Consideration of communication costs & \xmark       & \xmark        & \cmark        & \xmark       & \xmark     & \xmark         & \xmark     & \cmark    \\
		Optimization of random number generation  & \cmark       & \xmark        & \cmark        & \xmark       & \cmark     & \cmark         & \cmark     & \cmark    \\
		Reporting power consumption  & \cmark       & \xmark        & \cmark        & \cmark       & \cmark     & \xmark         & \xmark     & \cmark    \\
		\bottomrule
	\end{tabular}%
    \begin{tablenotes}
       \item [1] Accelerating on FPGA platforms.
       \item [2] Accelerating on ASIC platforms.
     \end{tablenotes}
    \end{threeparttable}
    \vspace{-5pt}
\end{table*}

\section{System Model and Security Assumptions}
This section defines the foundation of our work by establishing the implementation baseline and security model for \textsf{TFHE-SBC}. 
We present our choice of reference implementation and outline the security assumptions that guide our design approach.

\subsection{Baseline Implementation}
Among libraries supporting the TFHE scheme, TFHEpp~\cite{TFHEpp} offers more comprehensive functionality than alternatives such as OpenFHE~\cite{OpenFHE}~\cite{guimaraes2022mosfhet} and demonstrates superior performance compared to the original TFHE implementation~\cite{TFHE}. As our work represents the first TFHE framework optimized for single board computers, we adopt TFHEpp as our baseline for comparative evaluation.

\subsection{Threat Model and Security Assumptions}
Our threat model distinguishes between trusted and untrusted parties: the client operates in a trusted environment capable of key generation, encryption, and decryption, while the server represents an untrusted entity for computation offloading. 

We assume TFHE secret keys can be securely transferred to client devices, either before deployment or via secure channels. Deployed IoT devices must implement standard network protection mechanisms for confidentiality, integrity, and freshness of transmitted data. These mechanisms remain necessary even with homomorphically encrypted data, as they protect against active attacks and secure entire network messages. In our implementation, communication channels are protected using TLS.

HE provides protection against cloud-based adversaries, as no party can extract information from encrypted ciphertexts alone. While HE itself doesn't protect against malicious cloud attackers who might manipulate ciphertexts, it still preserves data privacy.

We consider direct attacks on client devices largely out of scope, as attackers with device access could directly access the underlying private data or data sources, making additional protections superfluous.

However, if secret keys are stored on devices, a successful attack could enable decryption of previously encrypted cloud data. 
To mitigate this risk, users can periodically update their device secret keys, limiting the window of vulnerability for historical data.
While this prevents joint analysis of data encrypted under different keys, it poses no significant limitations for private inference applications~\cite{benamira2023tt} that don't persistently store ciphertexts.

Finally, the post-quantum characteristics of HE provide inherent robustness against quantum attacks. Although our implementation doesn't specifically address quantum threats, quantum-secure communication can be achieved through existing libraries~\cite{libpqcrypto,wolfSSL} if required.

\section{TFHE Performance Analysis and Challenges}
\label{sec:bottleneck_tlwe_trlwe}
This section establishes the foundation for \textsf{TFHE-SBC} through two key analyses: first, a comparative assessment of TLWE versus TRLWE efficiency metrics, and second, an identification of critical performance bottlenecks in baseline implementations.
Our implementation parameters satisfy 128-bit security as specified in Table \ref{tab:param}.

\subsection{Comparative Analysis of TLWE and TRLWE}
\label{sec:tlwe}
Before conducting a detailed bottleneck analysis of encryption operations, we first compare TLWE and TRLWE to determine their relative efficiency in terms of computational and communication costs. This comparative analysis provides the foundation for our subsequent optimization strategies.

\noindent\textbf{Computation costs.}
TLWE encrypts plaintext per 1-bit and includes $N$ times uniformly distributed random number generation, the inner product of a vector of length $N$, and one time normally distributed random number generation.
TRLWE encryption involves $N$ bit encryption, $N$ times uniformly distributed random number generation, $N$ length polynomial multiplication, and $N$ times normally distributed random number generation.
Therefore, the time complexity of TLWE encryption is $\mathcal{O}(N)$ per bit, and that of TRLWE encryption is $\mathcal{O}(\log N)$ per bit since the time complexity of polynomial multiplication using FFT is $\mathcal{O}(N\log N)$.

\noindent\textbf{Communication costs.}
From $(a_0, \ldots, a_{N-1}, b) \in LWE_{\vec{s}, \sigma}(\Delta m) \subseteq \mathbb{Z}_q^{N+1}$, the TLWE ciphertext requires an array with length $N+1$ per bit.
From $(A, B) \in RLWE_{{S}, \sigma}(\Delta M) \subseteq \mathcal{R}_q^2$, the TRLWE ciphertext requires an array of length $2N$ per $N$ bit, so the size of the TLWE ciphertext is $\frac{N+1}{2}$ times larger than that of TRLWE ciphertext.
    \label{sec:ctxt_size}

\begin{table*}[t]
\caption{Parameters for 128-bit security in the TFHE scheme.}
    \centering
    \scriptsize
    \begin{tabular}{ccl}
    \toprule
    $q$ & $2^{32}$&The modulus for discretizing Torus.\\
    $N$ & $2^{10}$ &The length of the TLWE ciphertext. The dimension of TRLWE ciphertext. \\
    $\sigma$ & $2^{-25}$ &The standard deviation of the noise for the fresh TLWE and TRLWE ciphertext.\\
    \bottomrule
    \end{tabular}
    \label{tab:param}
    \vspace{-5pt}
\end{table*}

\subsection{Bottlenecks in TRLWE Implementation}
\label{sec:bottleneck}

We have demonstrated that TRLWE offers significant advantages over TLWE in terms of both computational efficiency and communication costs. 
In this section, we analyze the key bottlenecks in TRLWE encryption to inform the implementation in \textsf{TFHE-SBC} framework.

Table \ref{tab:rlwe_bottoleneck} shows the breakdown of TRLWE encryption on the Raspberry Pi Zero 2W using \texttt{trlweSymEncrypt} function in TFHEpp.
``Uniform Sampling'' and ``Gaussian Sampling'' refer to random number generation from their respective distributions, and ``PolyMul'' refers to polynomial multiplication. 
The most time-consuming operation is the generation of Gaussian random numbers, which account for 83.1\% of the total execution time.
Since random numbers sampled from the uniform and Gaussian distribution are generated the same number of times, random numbers sampled from the Gaussian distribution are generated slower than those sampled from the uniform distribution per random number generation.
PolyMul has less impact on encryption time than random number generation.
We give details about the bottleneck of each component and further discuss their optimization.


\noindent\textbf{Hardware dependent PRNG.}
\label{sec:csprng}
TFHEpp can use \texttt{/dev/urandom} provided by the Linux OS or Randen\footnote{\url{https://github.com/google/ randen}} as a cryptographically secure pseudorandom number generator (CSPRNG).
Randen is an implementation that requires an AES accelerator, but at this time, AES accelerators are not supported except for Raspberry Pi 5, so Table \ref{tab:rlwe_bottoleneck} is the execution time when using \texttt{/dev/urandom}. 
Random numbers obtained from CSPRNG are used to generate uniform and Gaussian random numbers, so the speed of pseudo-random number generation affects the speed of sampling from uniform and Gaussian distribution.
Moreover, it has been reported that random number generation using \texttt{/dev/urandom} slows down the encryption speed~\cite{banno2022oblivious}.
\cite{banno2022oblivious} pointed out that TRGSW encryption is 35 times faster on the SBC, which supports AES accelerators, than on the Raspberry Pi 4.
It is necessary to implement a CSPRNG that is fast even in environments where AES accelerators are not supported like the Raspberry Pi Zero series.

\noindent\textbf{Gaussian random number generator.}
\label{sec:polar}
The Gaussian random number generation algorithms include Box-Muller method~\cite{box1958note}, polar method~\cite{marsaglia1964convenient}, Kinderman method~\cite{kinderman1977computer}, Monty-Python method~\cite{marsaglia1998monty}, and Ziggurat method~\cite{marsaglia2000ziggurat}. 
In TFHEpp, Gaussian random numbers are generated by the \texttt{std::normal\_distribution} function included in the C++ standard library, and the random number generation algorithm is the polar method.
The polar method eliminates the trigonometric calculations required by the Box-Muller method but requires resampling with a probability of about 21.5\%.
This means that the expected number of uniformly distributed random numbers generated for generating a single Gaussian random number is 1.27.
The Ziggurat method, which has a lower rejection probability than the polar method, can speed up the process.

\begin{table}[t]
\caption{Breakdown of \texttt{trlweSymEncrypt} function in TFHEpp. Most of the non-optimized TRLWE encryption time consists of random number generation.}
    \centering
    \begin{tabular}{|c|c|c||c|}
    \hline
        \textbf{\;Uniform Sampling\;}&  \textbf{\;Gaussian Sampling\;} & \textbf{\;PolyMul\;} & \textbf{\;Total\;}\\\hline\hline
        7.7 [ms] & 38.3 [ms] & 0.16 [ms] &46.1 [ms] \\\hline
        16.7 \% & 83.1 \%& 0.35 \% & 100 \% \\\hline
    \end{tabular}
    \label{tab:rlwe_bottoleneck}
    \vspace{-5pt}
\end{table}

\subsection{Applicability of GPUs to Polynomial Multiplication}
For the polynomial multiplication, we enabled the option in TFHEpp to use the implementation for AArch64 of the FFT library SPQLIOS\footnote{\url{https://github.com/tfhe/experimental-tfhe/tree/master/circuit-Bootstrapping/src/ spqlios}}.
SPQLIOS does not use a real FFT, but rather packs the imaginary part of the input as a complex number, so that a polynomial of length $N$ can be transformed using an FFT with a polynomial of length $\frac{N}{2}$ as input.
We observed that SPQLIOUS optimizes well for CPU implementations, but we also investigate if the Raspberry Pi's GPU could enable faster polynomial multiplication.

Broadcom VideoCore IV\footnote{\url{https://docs.broadcom.com/doc/12358545}} is the GPU on the Raspberry Pi, giving a peak performance of 24GFlops.
Quad Processing Unit (QPU), the main arithmetic unit of the VC4, has four floating-point arithmetic units and operates as a 16-parallel SIMD over four cycles.
VC4 uses the host memory and reads and writes by DMA.
Of the 512 MB of host memory, VC4 is allocated 128 MB by default.
We explore whether the FFT, a component of polynomial multiplication, can be accelerated by VC4.

\begin{table}[t]
    \centering
    \footnotesize
    \caption{FFT/IFFT execution and {PolyMul} accuracy on VC4.}
    \begin{tabular}{|c||c|c|c|}\hline
         & \textbf{\;FFT [ms]\;} & \textbf{\;IFFT [ms]\;} & \textbf{\;RRMSE\;}\\\hline\hline
         CPU  & 0.054&0.061& $4.28 \times 10^{-7}$ \\\hline
         VC4 (GPU) & 0.034 & 0.036 & 0.015  \\\hline
    \end{tabular}
    \label{tab:gpu_fft}
    \vspace{-5pt}
\end{table}

The FFT implementation GPU\_FFT \cite{gpu_fft} provided by Raspberry Pi. 
This provides all power-of-2 FFT lengths between $2^8$ and $2^{22}$ points inclusive.
GPU\_FFT uses a kernel character device known as the ``mailbox'' for communication between the ARM and the Videocore. 

\noindent
\textbf{Execution time.}
Ignoring the memory copy between the CPU and VC4, Table \ref{tab:gpu_fft} shows that the FFT and IFFT are executed 1.59 times to 1.69 times faster on the VC4 than on the CPU by SPQLIOS with $N=2^{10}$.
Copying buffer from CPU to GPU requires 0.0045 ms, and from GPU to CPU requires 0.00096 ms, with the same $N$.
The memory transport time between the CPU and GPU does not exceed the main process, and the GPU provides a speed-up effect.

\noindent
\textbf{Accuracy.}
 PolyMul by FFT is known to have a small error due to floating-point arithmetic, but FFT on VC4, which supports only single precision, has additional errors.
In the TFHE scheme, since the maximum number of coefficients for PolyMul before rounding to the modulus $q$ is $N*(2^{31}-1)$, the number of bits required for exact PolyMul is $31+\log_2 N=42$ bits, where $N=2^{10}$.
The CPU implementation can retain this maximum value because it can operate in double precision.
However, GPU\_FFT uses single-precision floats for data and twiddle factors, which does not achieve sufficient PolyMul accuracy. 
This is because the QPU in VC4 contains two single-precision floating-point units, one for addition and one for multiplication, each operating on a vector of four single-precision numbers.
We compare the relative root mean square error (RRMSE) between native PolyMul without FFT / IFFT and PolyMul with FFT / IFFT, defined as 
$\text{RRMSE}(X,Y) = {\sqrt{\frac{1}{n}\sum_{i=1}^n |X_i-Y_i|^2}}/{\sqrt{\sum_{i=1}^n |X_i|^2}}$.
The native implementation provides exact polynomial multiplication, but its computational cost is $\mathcal{O}(N^2)$.
From Table \ref{tab:gpu_fft}, PolyMul on the single-precision VC4, the RRMSE is closer to 1, which is more inaccurate than PolyMul on CPU.
Therefore, we execute PolyMul entirely on the CPU to preserve accuracy, although FFT/IFFT on VC4 is faster.

\section{TFHE-SBC}
\begin{figure}[t]
    \centering
\includegraphics[width=\linewidth]{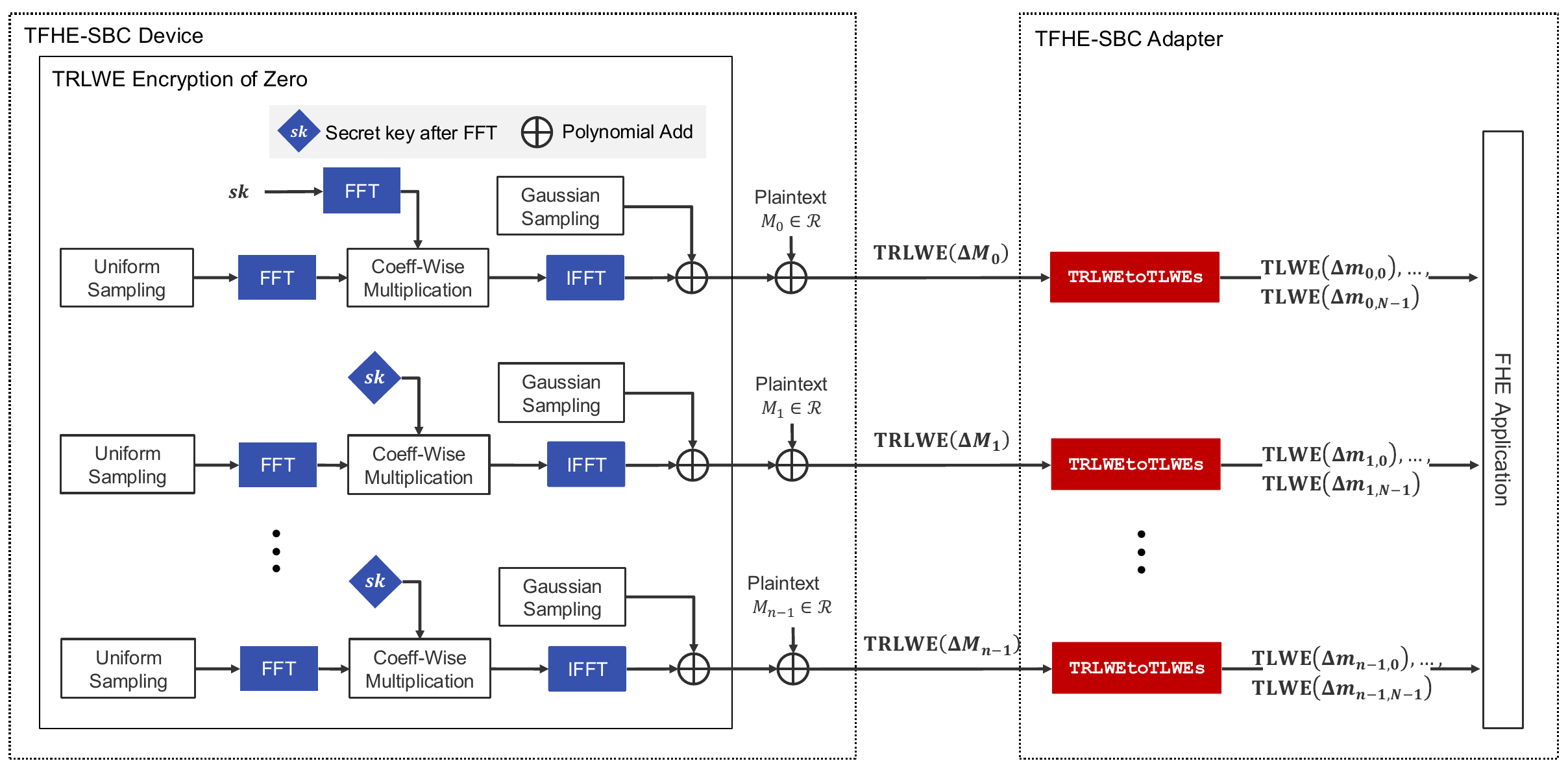}
    \caption{Architecture of \textsf{TFHE-SBC}: System workflow showing SBC-side TRLWE encryption and server-side conversion to TLWE ciphertext. The implementation features accelerated random number generation and single-execution FFT for secret keys.}
    \label{fig:tfhe-sbc}
    \vspace{-5pt}
\end{figure}

This section describes \textsf{TFHE-SBC}, a framework optimizing client-side TFHE operations for SBCs based on our bottleneck analysis in Section \ref{sec:bottleneck_tlwe_trlwe}. Our framework comprises two components: a client-side device library with optimized TRLWE encryption functions, and a server-side adapter library that converts encrypted data to TFHEpp-compatible format. Figure \ref{fig:tfhe-sbc} presents this architecture.

\subsection{TRLWE to TLWE Conversion Protocol}
\label{sec:faster_lwe}
TLWE encryption is significantly more computationally expensive and produces larger ciphertexts compared to TRLWE encryption as shown in Section \ref{sec:tlwe}.
To mitigate these inefficiencies at the client side, we propose to \textit{replace TLWE encryption with TRLWE encryption followed by Sample Extraction}~\cite{chillotti2020tfhe}.
As described in Section \ref{sec:sample-ext}, Sample Extraction converts a single TRLWE ciphertext into $N$ TLWE ciphertexts.
While Sample Extraction is traditionally a component used in the bootstrapping process, we repurpose it specifically to reduce client-side computational and communication costs.
In our protocol, the client transmits only the TRLWE ciphertext to the server, which then performs the conversion to TLWE ciphertext using Sample Extraction. This approach significantly reduces both the client-side execution time and ciphertext size.
Since Sample Extraction does not increase the noise, our protocol offers superior performance compared to Transciphering approaches~\cite{wei2023fregata,trama2023last,bon2023optimized} that require bootstrapping operations.
Our proposed protocol consists of the following steps:
\vspace{-3pt}
\begin{enumerate}
\item The trusted party generates the private key, Bootstrapping key, and key switching key, then securely shares the private key with the client device.
\item The client device \textit{encrypts the message using TRLWE} through our device library and shares the ciphertext with the cloud server. 
\item The cloud server \textit{converts the TRLWE ciphertext to TLWE ciphertext} using Sample Extraction via our adapter library.
\item The cloud server executes the application by homomorphic operations and shares the computation results with the trusted party.
\item The trusted party decrypts the result using the private key.
\end{enumerate}
\vspace{-3pt}

\vspace{-1em}
\subsection{Device Library}
\label{sec:devicelib}
The device library is structured to optimize performance on resource-constrained SBCs with particular focus on efficient random number generation and polynomial multiplication operations that address the critical bottlenecks identified in our analysis.

\noindent
\textbf{Pseudo-random number generation with BLAKE2.}
\label{sec:blake2}
In Section \ref{sec:csprng}, we identified that pseudo-random number generation by \texttt{/dev/urandom} is the bottleneck.
In \textsf{TFHE-SBC}, we mitigate this issue by obtaining only the random seed from \texttt{/dev/urandom} and expanding it using BLAKE2\footnote{\url{https://www.blake2.net/}}.
BLAKE2 is a cryptographic hash function that outperforms MD5, SHA-1, SHA-2, and SHA-3 in terms of speed, while maintaining security levels at least equivalent to the latest standard SHA-3.
This approach is also employed in the OpenFHE library for similar efficiency gains.

\noindent
\textbf{Gaussian random number with Ziggurat method.}
\label{sec:ziggurat}
In Section \ref{sec:polar}, we demonstrated that the generation of Gaussian random numbers using the polar method creates a bottleneck in TRLWE encryption within TFHEpp.
In \textsf{TFHE-SBC}, we implement the Ziggurat method for Gaussian random number generation, which offers superior performance compared to the polar method.
The Ziggurat method~\cite{marsaglia2000ziggurat} is considered the most efficient among the five algorithms discussed in Section \ref{sec:polar}.
As a rejection sampling algorithm, it relies on an underlying source of uniformly distributed random numbers and precomputed tables.
With a probability of 98.8\%, random numbers can be generated using only two floating-point multiplications. The expected number of uniformly distributed random numbers required to generate one Gaussian random number is 1.01, which is lower than that of the polar method.
In \textsf{TFHE-SBC}, we leverage the \texttt{boost::random::normal\_distribution} function from the Boost library\footnote{\url{https://www.boost.org/}} to implement the Ziggurat method efficiently.

\noindent
\textbf{Memory reuse in PolyMul.}
In the TFHEpp implementation of TRLWE encryption described in Section \ref{sec:bottleneck}, polynomial multiplication accounted for only 0.35\% of the execution time. However, after applying our random number generation optimizations, polynomial multiplication represents a larger percentage of the overall execution time.
Given that GPU-accelerated FFT proved ineffective for our use case, we focused on reducing the computational cost of polynomial multiplication on the CPU.
While TFHEpp already utilizes the refined FFT implementation from SPQLIOS, \textsf{TFHE-SBC} further reduces computation costs by reusing the Fourier-transformed secret key across multiple encryption operations.
Specifically, when the encryption function is invoked multiple times on an SBC, the uniform noise must be resampled and Fourier-transformed for each operation, but the secret key is transformed only once and subsequently reused.
This optimization strategy is illustrated in Figure \ref{fig:tfhe-sbc}.


\begin{wrapfigure}{R}{0.5\textwidth}
\vspace{-5em}
    \begin{minipage}{0.5\textwidth}
      \begin{algorithm}[H]
        \caption{Sample Extraction: \texttt{TRLWEtoTLWEs}~\cite{chillotti2020tfhe}}
        \label{alg:convert}
        \begin{algorithmic}
        \REQUIRE $(A, B) \in \text{TRLWE}(\Delta M)$
        \ENSURE $(a_{0,0}, \ldots, a_{0,N-1}, b_0) \in \text{TLWE}(\Delta m_0)$, $\ldots$, $(a_{N-1,0}, \ldots, a_{N-1,N-1}, b_{N-1}) \in \text{TLWE}(\Delta m_{N-1})$
        \FOR{$0 \leq h < N$}
        \FOR{$0 \leq i \leq h$}
            \STATE $a_{h,i} = A[h - i];$
        \ENDFOR
        \FOR{$h + 1 \leq i < N$}
            \STATE $a_{h,i} = - A[N + h - i];$
        \ENDFOR
            \STATE $b_h = B[h];$
        \ENDFOR
        \RETURN \small{$(a_{0,0}, \ldots, a_{0,N-1}, b_0),$ $\ldots,$ $(a_{N-1,0}, \ldots, a_{N-1,N-1}, b_{N-1})$}
        \end{algorithmic}
      \end{algorithm}
    \end{minipage}
    \vspace{-2em}
\end{wrapfigure}

\vspace{-1.0em}
\leavevmode
\subsection{Adapter Library}
We additionally provide an ``adapter'' server module to convert data encrypted by \textsf{TFHE-SBC} to be compatible with the TFHEpp library.
The adapter library converts TRLWE ciphertexts received from the client into TLWE ciphertexts that can be handled by TFHEpp, as detailed in Algorithm \ref{alg:convert}.
The conversion process leverages the \texttt{SampleExtractIndex} function from TFHEpp, which selectively copies elements from the TRLWE ciphertext array to construct the TLWE ciphertext.
When an SBC inputs a TRLWE ciphertext $\text{TRLWE}(\Delta M)$ where $M = \sum_{j=0}^{N-1} m_j X^j \in \mathcal{R}_p$, the adapter library generates the corresponding TLWE ciphertexts $\text{TLWE}(\Delta m_0), \text{TLWE}(\Delta m_1)$, $\ldots, \text{TLWE}(\Delta m_{N-1})$.
The resulting ciphertexts can be seamlessly integrated with any application implemented using the TFHEpp library.

\section{Experiments}
As established in Section 1, our primary objectives are to optimize client-side TFHE operations and reduce communication costs.
This section addresses the following research questions through comprehensive empirical evaluations to quantify the extent to which our solution meets these requirements:

\begin{enumerate}
    \item[\textbf{Q1}:] How does \textsf{TFHE-SBC} improve computational efficiency on both the SBC client and server?
    \item[\textbf{Q2}:] How does \textsf{TFHE-SBC} reduce the communication overhead between the SBC and the server?
    \item[\textbf{Q3}:] How does the energy efficiency of \textsf{TFHE-SBC}'s client-side operations compare to alternative methods?
\end{enumerate}

For our experiments, we use the parameter values for TLWE and TRLWE as specified in Table \ref{tab:param}, which correspond to the \texttt{lvl1param} settings in the TFHEpp library.
These parameters achieve 128-bit security.
All experiments are conducted on a Raspberry Pi Zero 2W (ARM 1GHz 4-core CPU, 512MB RAM) as the SBC client and a server machine (Intel 1.9GHz 56-core CPU, 512GB RAM) as the cloud server.
We implemented \textsf{TFHE-SBC} in C++17 and compiled it with GCC-10.2.1.
Our implementation is available at \url{https://anonymous.4open.science/r/TFHE-SBC-A178}.

To address \textbf{Q1} and \textbf{Q2}, we evaluate and compare the encryption time on the SBC and ciphertext size for both TLWE and TRLWE encryption methods. We also measure the conversion time from TRLWE ciphertext to TLWE ciphertext on the server.
Additionally, we analyze the execution time of TRLWE encryption across different random number generation and polynomial multiplication methods to determine the optimal configuration for \textsf{TFHE-SBC}.
We report memory utilization metrics for each configuration to provide a comprehensive performance assessment.

For \textbf{Q3}, we employ a UM24 USB power meter connected between the power supply and the device to precisely measure energy consumption during client-side operations.

\begin{table}[t]
    \centering
    \caption{Runtime and data memory requirements for each noise sampling.}
    \begin{tabular}{|c|c||c||c|}\hline
   {Sampling} &{Configuration}& \textbf{Runtime [ms]} & \textbf{RAM [KiB]}\\\hline\hline
        \multirow{2}{*}{Uniform}&\texttt{/dev/urandom}&7.71 &192.1\\\cline{2-4}
                                        &BLAKE2 & 0.92 &196.2\\\hline\hline
        \multirow{8}{*}{Gaussian}&\texttt{/dev/urandom}& \multirow{2}{*}{38.26}&\multirow{2}{*}{192.1}\\
        &Polar&&\\\cline{2-4}
        &\texttt{/dev/urandom}& \multirow{2}{*}{15.36}&\multirow{2}{*}{192.2}\\
        &Ziggurat&&\\\cline{2-4}
        &{BLAKE2}& \multirow{2}{*}{4.98}&\multirow{2}{*}{196.2}\\
        &Polar&&\\\cline{2-4}
        &{BLAKE2}& \multirow{2}{*}{1.98}&\multirow{2}{*}{196.2}\\
        &Ziggurat&&\\\cline{2-4}
        \hline
    \end{tabular}
    \label{tab:sampling}
    \vspace{-5pt}
\end{table}

\begin{table}[t]
    \centering
    \caption{Breakdown of PolyMul}
    \begin{tabular}{|c||c|c|c||c|}\hline
   \multirow{2}{*}{Configuration} &\multicolumn{3}{c||}{\textbf{Runtime [ms]}} & \multirow{2}{*}{\textbf{RAM [KiB]}}\\\cline{2-4}
   &FFT&Coeff-Wise Mult&IFFT&\\\hline\hline
       {on-the-fly}&0.103&0.0086 &0.0611&314.6\\\hline
     {reuse}&0.065&0.0086&0.0611&314.6\\
        \hline
    \end{tabular}
    \label{tab:polymul}
    \vspace{-5pt}
\end{table}

\subsection{Experimental Results}
To {address} \textbf{Q1}, we first evaluate the computational {efficiency} on the SBC side and then on the server side.
We {further analyze specific} optimization effects in TRLWE encryption for uniform and Gaussian noise generation {as well as} polynomial multiplication.

\noindent
\textbf{End-to-end {computational efficiency} on SBCs.}
Table \ref{tab:ex-tlwe} {presents a comparison of} execution time, ciphertext size, and energy consumption {across} TLWE encryption (\texttt{tlweSymEncrypt} function) and TRLWE encryption (\texttt{{trlweSymEncrypt}} function) by the TFHEpp library as {baselines}, and \textsf{TFHE-SBC}.
\textsf{TFHE-SBC} {measurements represent} the {optimal} combination of random number generation and polynomial multiplication as described in Section \ref{sec:devicelib}.
TFHEpp in environments without AES accelerators uses \texttt{/dev/urandom} as CSPRNG and the polar method for Gaussian noise.
{Regarding} execution time, {since} TRLWE encryption {processes} $N$ bits at a time, the encryption times for 256-bit and 1024-bit {plaintexts} are the same.
The optimized TRLWE with \textsf{TFHE-SBC} is approximately 15 times faster than TRLWE encryption with TFHEpp library and approximately 2490 times faster than TLWE encryption.
For RAM consumption, \textsf{TFHE-SBC} is 12.8 times more efficient for 1024-bit encryption than baseline TLWE encryption.
For TRLWE, memory usage was 336.2 KiB for our method, 1.01 times the baseline.
The 4 KiB increase in memory usage in \textsf{TFHE-SBC} is due to the random number generation being replaced by BLAKE2, as {evidenced in} Table \ref{tab:sampling}.
This {demonstrates} that \textsf{TFHE-SBC} achieved 15 times faster encryption without much increase in memory usage from non-optimized TRLWE.

\begin{wrapfigure}{r}{0.5\textwidth}
\vspace{-2em}
    \centering
    \begin{subfigure}[b]{0.23\columnwidth}
         \centering
         \includegraphics[width=\textwidth]{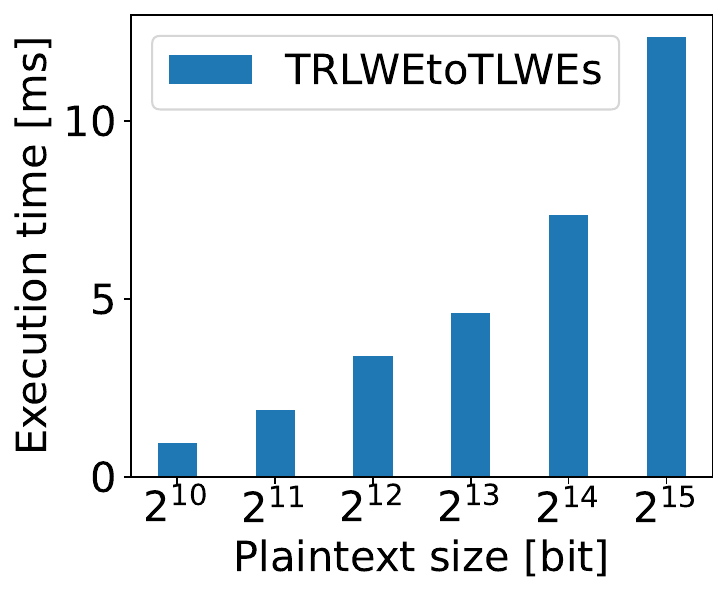}
         \caption{Execution time}
         \label{fig:switch_time}
     \end{subfigure}
     \hfill
     \begin{subfigure}[b]{0.24\columnwidth}
         \centering
         \includegraphics[width=\textwidth]{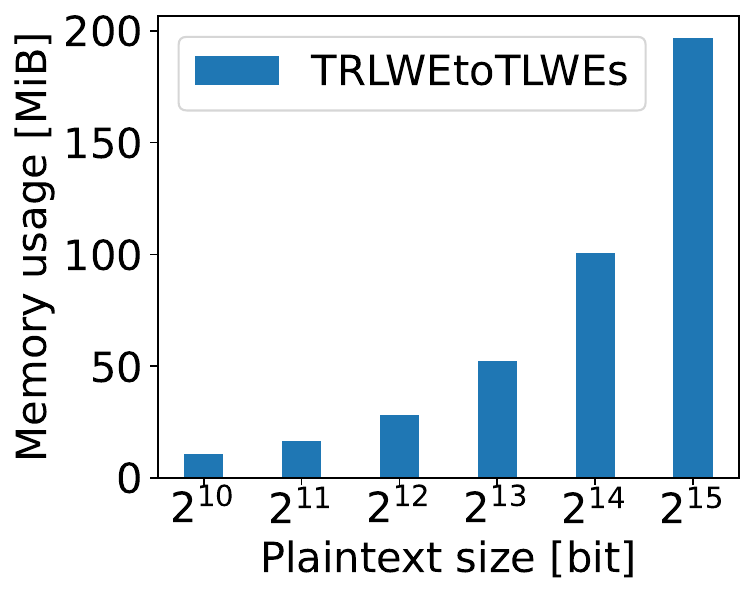}
         \caption{Memory usage}
         \label{fig:server_mem}
     \end{subfigure}
          \caption{Performance of \texttt{TRLWEtoTLWEs}. {Results confirm} \textsf{TFHE-SBC}'s end-to-end encryption outperforms direct TLWE encryption, even with conversion overhead.}
     \label{fig:eval_server}
     \vspace{-2em}
\end{wrapfigure}

\noindent
\textbf{{TRLWE to TLWE conversion performance}.}
We evaluate the server-side computational {overhead of} the \textsf{TFHE-SBC} adapter library. Note that if the client is TLWE encrypted, such costs are ignored.
Figure \ref{fig:switch_time} shows the execution time of \texttt{TRLWEtoTLWEs} on the server side with \textsf{TFHE-SBC}'s adapter library.
The execution is single-threaded.
For a 1024-bit plaintext, TRLWE encryption with the \textsf{TFHE-SBC}'s device library requires 3.1 ms on the client device and 1.0 ms for conversion on the server, for a total of 4.1 ms.
In this case, TLWE encryption requires 7705.6 ms from Table \ref{tab:ex-tlwe}.
Therefore, end-to-end encryption of \textsf{TFHE-SBC} is faster than TLWE encryption on the client, even when the total execution time of \texttt{TRLWEtoTLWEs} is taken into account.
Figure \ref{fig:server_mem} shows that the peak memory usage of \texttt{TRLWEtoTLWE} increases linearly.
\texttt{TRLWEtoTLWE} requires 8KiB TRLWE ciphertext as input and outputs 4MiB TLWE ciphertexts, so at least $4*\frac{\text{plaintext size}}{N}$ MiB is reserved. 
Additional memory is required to concatenate the output TLWE by \texttt{std::copy}.

\noindent
\textbf{{Noise sampling optimization analysis}.}
From Table \ref{tab:sampling}, changing from \texttt{/dev/urandom} to BLAKE2 as the CSPRNG had the greatest speedup effect.
Since CSPRNG is involved in generating random numbers sampled from uniform and Gaussian distributions, {enhancing pseudorandom number generation directly improves sampling performance}.
BLAKE2 generates random numbers 8 times faster than \texttt{/dev/urandom} for both distributions with an additional 4 KiB increase in memory usage.
As {detailed in} Section \ref{sec:ziggurat} and Table \ref{tab:sampling}, the {Ziggurat} method {delivers substantial improvements over} the polar method employed in TFHEpp.
When implemented with the same CSPRNG, Gaussian sampling using the {Ziggurat} method is more than twice as fast as the polar method.
{This performance differential stems from} the difference in rejection probabilities of each method.

\noindent
\textbf{Polynomial multiplication {optimization}.}
We test the performance of PolyMul with two configurations, as shown in Table \ref{tab:polymul}.
In the {``on-the-fly''} version, we execute FFTs for the secret key and the uniform noise, so the FFT is performed twice for each encryption.
In the {``reuse''} flavor, the FFT value of the secret key is precomputed.
It requires less computation than the {``on-the-fly''} case since the FFT for the secret key only needs to be calculated once per encryption.
Peak memory usage is the same for both cases.
PolyMul {``reuse''} case, which eliminates one FFT execution, improves the computational cost of the PolyMul {``on-the-fly''} case by 28\%.

To {address} \textbf{Q2}, Table \ref{tab:ex-tlwe} shows the communication costs between SBCs and the server.
The communication cost here refers to the ciphertext size.
The ciphertext size of TLWE, which encrypts one bit at a time, is approximately 512 times larger than that of TRLWE, which encrypts $N$ bits at once.
This result is consistent with the theoretical result in Section \ref{sec:ctxt_size}.
{These findings confirm} that generating TRLWE ciphertext leads to a reduction in the amount of client-server communication.

To {address} \textbf{Q3}, we conduct measurements of the SBC's power consumption.
The rightmost column of Table \ref{tab:ex-tlwe} shows the energy consumed in the end-to-end encryption for each plaintext size on SBC.
The idle power consumption in Raspberry Pi Zero 2W is 600 mW.
\textsf{TFHE-SBC} consumes 5 mJ per encryption, nearly 2000 times less energy than TLWE encryption.
In addition, our approach achieves 12 times more energy efficiency than baseline TRLWE encryption.
This {substantial improvement directly correlates with} \textsf{TFHE-SBC}'s {enhanced} encryption runtime.


\section{Conclusion and Future Work}
We presented \textsf{TFHE-SBC}, the first implementation of TFHE client-side operations designed specifically for resource-constrained SBCs. 
Our framework achieves significant improvements in both computational efficiency and communication overhead by employing TRLWE encryption on SBCs with server-side conversion to TLWE ciphertext. We developed key optimizations tailored to SBCs' limited resources, including enhanced random number generation and efficient polynomial multiplication with memory reuse. Our investigation of GPU acceleration on the Raspberry Pi identified critical accuracy limitations that informed our CPU-focused design.

Experimental evaluations demonstrate that \textsf{TFHE-SBC} outperforms state-of-the-art implementations on the Raspberry Pi Zero 2W across all metrics: achieving encryption speeds 15 to 2486 times faster, communication efficiency 512 times greater, memory utilization up to 12.8 times more efficient, and energy consumption 12 to 2004 times lower than existing approaches.

While \textsf{TFHE-SBC} makes TFHE more practical for IoT applications, our current implementation supports only private key encryption. Extending the framework to support public key encryption presents additional challenges, as it requires substantially more noise sampling operations and memory resources. Addressing these challenges remains an important direction for future research.

\bibliographystyle{unsrt}  
\bibliography{references}
\end{document}